\documentclass[prd,twocolumn,floatfix,preprintnumbers,letterpaper]{revtex4}

\usepackage{graphicx}
\input{epsf}
\usepackage{epsf}
\usepackage{graphicx,epsfig}
\usepackage{bm}
\usepackage{latexsym,amssymb,amsmath,float}

\newcommand {\ga} {\ {\raise-.5ex\hbox{$\buildrel>\over\sim$}}\ }
\newcommand {\la} {\ {\raise-.5ex\hbox{$\buildrel<\over\sim$}}\ }

\begin{document}

\title{Limits on MeV Dark Matter from the Effective Number of Neutrinos}
\author{Chiu Man Ho and Robert J. Scherrer}
\affiliation{Department of Physics and Astronomy, Vanderbilt University,
Nashville, TN  ~~37235}

\begin{abstract}
Thermal dark matter that couples
more strongly to electrons and photons than to neutrinos will heat the electron-photon
plasma relative to the neutrino
background if it becomes nonrelativistic
after the neutrinos decouple from the thermal background.  This results
in a reduction in $N_{eff}$ below the standard-model value, a result strongly
disfavored by current CMB observations.  Taking conservative lower bounds
on $N_{eff}$ and on the decoupling temperature of the neutrinos, we derive a
bound on the dark matter particle mass of $m_\chi > 3-9$ MeV, depending on the spin and
statistics of the particle.  For $p$-wave annihilation, our limit
on the dark matter particle mass is stronger than the limit derived
from distortions to the CMB fluctuation spectrum produced by annihilations
near the epoch of recombination.
\end{abstract}

\maketitle

Roughly $20-25$\% of the total energy content of the universe is in the form of
non-baryonic dark matter.  While a dark matter particle mass
in the GeV range is often assumed, there has also been interest in masses
in the MeV range.  Dark matter with a mass in this range was invoked to
explain the 511 keV $\gamma$-rays observed by INTEGRAL \cite{Boehm}, and to
explain the cosmic $\gamma$-ray background at $1-20$ MeV \cite{Ahn}.
Supersymmetric models with MeV dark matter have been proposed \cite{Hooper},
and MeV dark matter can arise in the context of the WIMPless dark matter model
\cite{Feng}. MeV dark matter can have interesting effects on large-scale structure
\cite{Kaplinghat}.

We note here that a thermal MeV dark matter particle that couples more strongly to electrons and photons
than to neutrinos will heat the electron-photon plasma when it becomes nonrelativistic
before its abundance freezes out. If this occurs after the neutrinos decouple from the thermal background,
then the ratio of the neutrino temperature to the photon temperature
will be reduced, a process similar to the heating that occurs when the
electron-positron pairs become nonrelativistic.  The final result is a decrease
in the effective number of neutrino degrees of freedom.  This effect
was first explored by Kolb et al. \cite{KTW} and more recently by Serpico and Raffelt
\cite{SR} in the context of primordial nucleosynthesis.  Recent
CMB observations \cite{WMAP7B,Dunkley,Keisler} place severe lower bounds
on $N_{eff}$, allowing us to constrain this process.  (See also the earlier work of
Ref. \cite{Dorosh}, which examined heating of the photons relative to the neutrinos
from decaying particles).

At recombination, the energy density in relativistic particles includes photons, whose temperature,
$T_\gamma$, and therefore energy density is extremely well-measured, and a neutrino background with temperature
$T_\nu = (4/11)^{1/3} T_\gamma$. The theoretical prediction for the effective number of neutrinos (assuming slight reheating
of the neutrinos from early $e^+e^-$ annihilation) is $N_{eff} = 3.046$ \cite{Dolgov,mangano1}.  The
neutrino density cannot be measured directly, but it can be inferred from measurements of the CMB.
(For a discussion of the effect of $N_{eff}$ on the CMB
fluctuations, see Refs. \cite{Bashinsky,Hou}). The values of $N_{eff}$ from recent CMB observations, in combination with other
cosmological data, are
$N_{eff} = 4.34^{+0.86}_{-0.88}$ (68\% CL) from WMAP
\cite{WMAP7B},
$N_{eff} = 4.56 \pm 0.75$ (68\% CL) from the Atacama Cosmology Telescope \cite{Dunkley},
and $N_{eff} = 3.86 \pm 0.42$ (68\% CL) from the South Pole Telescope \cite{Keisler}.
Archidiacono et al. \cite{Arch} used combined datasets to derive
$N_{eff} = 4.08^{+0.71}_{-0.68}$ (95\% CL).  Clearly, the data favor values
of $N_{eff}$ larger than the standard-model theoretical prediction, rather
than smaller.

The extent of the heating from dark matter annihilation in the early universe
can be derived from entropy conservation (see Refs. \cite{Weinberg,KT}, from which
our discussion is derived).  Our paper assumes a dark matter particle that couples much more strongly
to electrons and photons than to neutrinos. The most natural example of such a particle is one that
interacts with ordinary matter through an electromagnetic form factor, such as an electric or magnetic dipole \cite{Pospelov,Sigurdson,Gardner,Masso,Fitzpatrick,Cho,Heo1,Heo2,Banks,Barger1,Fortin,
Nobile,Barger2,Heo3}, or an anapole moment \cite{HoScherrer2}. Dark matter particles in this category annihilate into Standard
Model particles through the mediation of photons, while the models considered by
Refs. \cite{Boehm,Ahn,Hooper,Feng,Kaplinghat} require the mediation of a new fermion or vector boson.
In fact, the dark matter particles considered in Refs. \cite{Boehm,Ahn,Hooper,Feng} could be relevant if their coupling
with neutrinos is postulated to be suppressed. However, the model considered by Ref. \cite{Kaplinghat} requires that the dark
matter particle couples to electrons and neutrinos equally, and so it is not relevant.

Let $\chi \bar \chi$ denote the pair of dark matter particles. To make our study general,
we will allow a range of possibilities for the dark matter, including
a self-conjugate scalar, a non-self-conjugate scalar, a spin-1/2 Majorana fermion or a spin-1/2 Dirac fermion.
Thus, for the cases with self-conjugate and non-self-conjugate scalars, the notation $\chi \bar \chi$ really means $\chi \chi$
and $\chi \chi^\ast$ respectively. But for simplicity, we will keep the notation $\chi \bar \chi$ throughout the paper.

Consider first the case where the dark matter annihilates entirely after the neutrinos
decouple, which occurs at a temperature of $T_{d} \approx 2-3$ MeV \cite{Dolgov,Enqvist}.
The total entropy prior to $\chi \bar \chi$ annihilation is proportional to
\begin{equation}
S = \frac{R^3}{T}(\rho_{e^+e^-} +  \rho_{\gamma} + \rho_{\chi \bar \chi} + p_{e^+e^-} +
p_{\gamma} + p_{\chi \bar \chi}),
\end{equation}
while after $\chi \bar \chi$ annihilation it is
\begin{equation}
S = \frac{R^3}{T}(\rho_{e^+e^-} +  \rho_{\gamma} + p_{e^+e^-} +
p_{\gamma}).
\end{equation}
For a relativistic particle, $p = \rho/3$, so following Ref. \cite{KT},
we can write the total entropy density as
\begin{equation}
s=\frac{\rho_{\textrm{tot}}+p_{\textrm{tot}}}{T} = \frac{2 \pi^2}{45} g_{*S} T^3,
\end{equation}
where $g_{*S}$ is the total number of spin degrees of freedom for bosons, and 7/8 times the total number
of spin degrees of freedom for fermions.  Then the total entropy is
\begin{equation}
S = \frac{2 \pi^2}{45} g_{*S} (RT)^3,
\end{equation}
which is conserved through the process of any particle becoming nonrelativistic and
annihilating. So the ratio of the final value of $RT$ after annihilation to the initial value of $RT$
prior to annihilation is
\begin{equation}
\label{heating}
\frac{(RT)_f}{(RT)_i} = \left(\frac{g_{*Si}}{g_{*Sf}}\right)^{1/3},
\end{equation}
where $g_{*Si}$ and $g_{*Sf}$ are the values of $g_{*S}$ for the relativistic
particles in thermal equilibrium before and after annihilation, respectively.  When the $\chi \bar \chi$ pairs
annihilate after neutrino decoupling, the neutrinos do not share in the heating, so that
$RT_\nu$ is constant and $T_\nu \propto R^{-1}$, while the photons and electron-positron pairs are heated as in
Eq. (\ref{heating}).
Therefore, for the $\chi \bar \chi$ pairs with $g$ internal degrees of freedom, the ratio of $T_\nu$ to $T_\gamma$ after $\chi \bar \chi$
annihilation is:
\begin{equation}
\label{ratf}
T_\nu/T_\gamma = \left[\frac{(7/8)4 + 2}{(7/8)4 + 2 + (7/8)g}\right]^{1/3},
\end{equation}
if $\chi$ is a fermion, and
\begin{equation}
\label{ratb}
T_\nu/T_\gamma = \left[\frac{(7/8)4 + 2}{(7/8)4 + 2 + g}\right]^{1/3},
\end{equation}
if it is a boson.
Taking, for example, the $\chi$ particle to be a spin-1/2 Majorana fermion gives $g = 2$,
so that $T_{\nu}/T_\gamma = (22/29)^{1/3}$.
Subsequent $e^+e^-$ annihilation further heats the photon temperature relative to the neutrino
temperature by a factor of $(11/4)^{1/3}$, so that the final ratio of the neutrino temperature to
the photon temperature would be $(88/319)^{1/3}$.

In terms of $N_{eff}$, the energy density for neutrinos is
given by
\begin{equation}
\rho_\nu = N_{eff}\left(\frac{7}{8}\right)\left(2\right)\left(\frac{\pi^2}{30}\right)
\left(\frac{T_\nu}{T_\gamma}\right)^4 T_\gamma^4.
\end{equation}
Since $\rho_\nu$ at fixed $T_\gamma$ is the quantity that is inferred from CMB observations,
a change in $T_\nu/T_\gamma$ will be interpreted as a change in $N_{eff}$, with $N_{eff} \propto
(T_\nu/T_\gamma)^4$.
In this case, $\chi \bar \chi$ annihilation reduces the value of $T_\nu/T_\gamma$
relative to its value in the standard model
by a factor of $(22/29)^{1/3}$,
which corresponds to $N_{eff} = 3(22/29)^{4/3} = 2.1$, a value clearly excluded by the CMB observations.

This value of $N_{eff}$ corresponds to a dark matter particle with a mass well below the neutrino
decoupling temperature.  However, to derive a useful limit, we must consider what happens
when $\chi$ annihilates during neutrino decoupling.  Neutrino decoupling is not a sudden process,
but for the purposes of our simplified calculation, we will take it to occur abruptly
at a fixed temperature $T_d$, and we will assume that dark matter annihilations before $T_d$
fully heat the neutrinos, while those after $T_d$ heat only the photons and $e^+ e^-$ pairs.
Let $I(T_\gamma)$ be given by (see, e.g., Ref. \cite{Weinberg} for a similar calculation):
\begin{widetext}
\begin{eqnarray}
I(T_\gamma)&\equiv& \frac{S}{(RT_\gamma)^3} = \frac{1}{T_\gamma^4}\,(\rho_{e^+e^-} +  \rho_{\gamma} + \rho_{\chi \bar \chi} + p_{e^+e^-} +
p_{\gamma} + p_{\chi \bar \chi}), \nonumber\\
\label{formula}
&=& \frac{11}{45} \pi^2 + \frac {g}{2 \pi^2} \int_{x=0}^\infty x^2 dx \left
(\sqrt{x^2 + (m_\chi/T_\gamma)^2} + \frac{x^2}{3\sqrt{x^2 + (m_\chi/T_\gamma)^2}}\right)
\left[\exp(\sqrt{x^2 + (m_\chi/T_\gamma)^2} \pm 1)\right]^{-1},
\end{eqnarray}
\end{widetext}
where the plus (minus) sign is for a fermionic (bosonic) dark matter particle, and
the variable of integration is $x = p_\chi/T_\gamma$.  In the limit where
all particles are fully relativistic, $I$ reduces to $(2 \pi^2/45)g_{*S}$; the integral
in Eq. (\ref{formula}) just quantifies the contribution to $I$ from
$\chi \bar \chi$ as they become nonrelativistic.

As mentioned above, the $\chi \bar \chi$ annihilation will heat up photons relative to neutrinos only after neutrino
decoupling. But this heating ends when the $\chi \bar \chi$ particles drop out of thermal equilibrium.
Thus, the ratio of the neutrino temperature to the photon temperature due
to $\chi \bar \chi$ annihilation alone is
\begin{equation}
T_\nu/T_\gamma = \left[\frac{I(T_f)}{I(T_d)}\right]^{1/3},
\end{equation}
where $T_f$ is the temperature at which the $\chi \bar \chi$ particles freeze out.
Since $m_\chi/T_f \sim 20$ \cite{KT}, it is obvious from Eq. (\ref{formula})
that we can simply set $T_f = 0$ with negligible error:
\begin{equation}
\label{Tratio}
T_\nu/T_\gamma = \left[\frac{I(0)}{I(T_d)}\right]^{1/3}.
\end{equation}
The physical reason for this is that the $\chi \bar \chi$ abundance
freezes out at a temperature of $T_f \sim m_\chi/20$, while most of the entropy
from the $\chi \bar \chi$ annihilations is transferred to the thermal
background when $T \sim m_\chi/3$.
Of course, the temperature ratio given by Eq. (\ref{Tratio}) must then be multiplied by an additional factor
of $(4/11)^{1/3}$ from $e^+ e^-$ annihilations to obtain the final ratio
of the neutrino temperature to the photon temperature.

In this approximation, the effective number of neutrinos as measured by CMB experiments will
be given by
\begin{equation}
N_{eff} = 3.046\left[\frac{I(0)}{I(T_d)}\right]^{4/3}.
\end{equation}
The value of $N_{eff}$ as a function of $m_\chi/T_d$
is shown in Fig. 1, for a self-conjugate scalar boson
($g=1$), a non-self-conjugate scalar boson
($g=2$), a spin-1/2 Majorana fermion
($g=2$) and a spin-1/2 Dirac fermion ($g=4$).
\begin{figure}[t!]
\centerline{\epsfxsize=3.8truein\epsffile{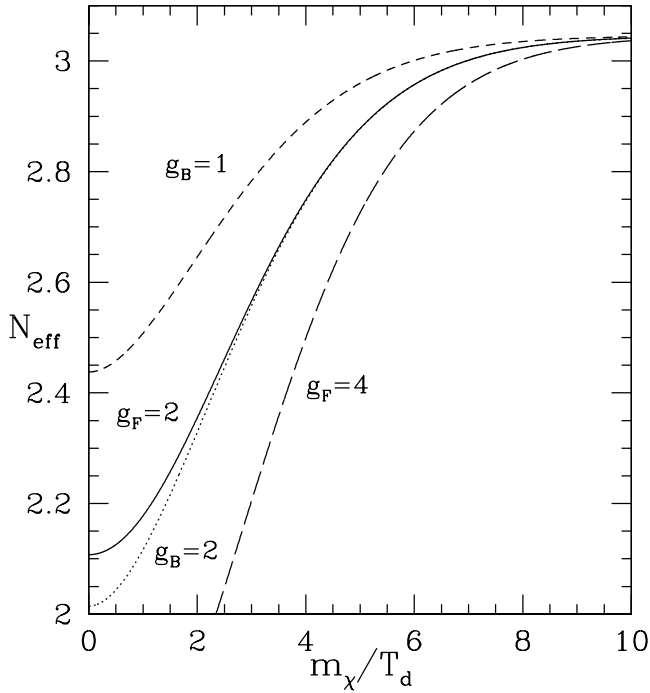}}
\caption{The effective number of neutrino degrees of freedom, $N_{eff}$,
that would be deduced from cosmic microwave background observations
for a thermal dark matter particle with mass $m_\chi$,
assuming sudden decoupling of the cosmic
neutrinos at a temperature $T_d$.  Curves correspond, top to bottom,
to a $g = 1$ boson (short dash), $g = 2$ fermion (solid), $g = 2$ boson (dotted), and
$g = 4$ fermion (long dash).}
\end{figure}

In fact, from Eqs. (\ref{ratf})-(\ref{ratb}), we can derive the $m_\chi \ll T_d$ limit
for $N_{eff}$, namely
\begin{equation}
N_{eff} = 3.046\left[\frac{11}{11 + (7/4)g}\right]^{4/3},
\end{equation}
for fermionic $\chi$, and
\begin{equation}
N_{eff} = 3.046\left[\frac{11}{11 + 2g}\right]^{4/3},
\end{equation}
for bosonic $\chi$.

As noted earlier, neutrino decoupling is not a sudden process, so $T_d$ is not
completely well-defined.  Ref. \cite{Enqvist} gives a widely cited value of
$T_d = 2.3$ MeV for the electron neutrinos, with the $\mu$ and $\tau$ neutrinos decoupling
at a higher temperature. However, neutrino oscillations will tend to equilibrate
the decoupling of all three neutrinos, an effect discussed in Refs.
\cite{mangano1,Hannestad}.  Here we will simply take $T_d \ga 2$ MeV as a conservative
lower bound. Note that the presence of the additional relativistic energy
density from the $\chi \bar \chi$ particles themselves will increase $T_d$, but this
turns out to be a miniscule effect \cite{HoScherrer3}.

Now we must determine a reasonable lower bound on $N_{eff}$.
The combined results from Refs. \cite{WMAP7B,Dunkley,Keisler} are barely consistent
with the standard model value of $N_{eff} = 3.046$.
However, we will err on the side of caution and choose
a lower bound of $N_{eff} > 2.6$, which is excluded at $2\sigma$ by all three
sets of CMB observations.

These limits on $N_{eff}$ and $T_d$ can be combined with the results displayed
in Fig. 1 to derive a lower bound on $m_\chi$.  These bounds are $m_\chi \ga 3$
MeV for the self-conjugate scalar boson, $m_\chi \ga 6$ MeV for a
two-component boson or fermion, and $m_\chi \ga 9$ MeV for a Dirac fermion.

These limits are relevant for several models in the literature.  As noted by
Beacom and Yuksel \cite{Beacom}, the model proposed in Ref. \cite{Boehm}
actually requires positron injection at very low energies ($\la 3$ MeV) to produce the
511 keV $\gamma$-rays observed by INTEGRAL \cite{Boehm}.
But dark matter masses low enough to produce such particles from annihilations
are ruled out by our limit.  Thermal dark matter with the correct relic abundance
interacting through an electric or magnetic dipole moment must have a mass less than $1-10$ GeV to
avoid conflict with direct detection experiments \cite{Fortin}; our results shrink the allowed
window from the other direction.

Our limits are complementary to several others in the literature.  As noted,
dark matter particles with masses in this range also affect primordial
nucleosynthesis, and bounds can be placed from the observed element abundances,
particularly helium-4.  However, the effect on $N_{eff}$ as measured by the
CMB appears to provide a better limit.  For example, in the $1-10$ MeV mass range,
Serpico and Raffelt \cite{SR} found a maximum reduction of only 0.002 in the
primordial helium mass fraction.   Using the results of Ref. \cite{Steigman},
this corresponds to $\Delta N_{eff} = -0.15$, much smaller than the typical
values in Fig. 1.  However, there is no contradiction between our results and those
of Ref. \cite{SR}.  When $T_\nu/T_\gamma$
is reduced prior to primordial nucleosynthesis, there are actually two
effects on the helium-4 abundance. First, the reduction in the expansion rate
at fixed $T_\gamma$ reduces the helium-4 abundance, and this is the dominant
effect, as noted by Serpico and Raffelt.  However, there is a second effect
which partially cancels the first:  the decrease in the electron neutrino
temperature reduces the weak interaction rates, which tends to increase
the helium-4 abundance.  Thus, the effect on BBN is smaller than if one
reduced the overall expansion rate alone.

Another lower bound on $m_\chi$ comes from distortions to the CMB fluctuation
spectrum due to annihilations near the epoch of recombination \cite{Padmanabhan,
Mapelli,Zhang,Galli,Hutsi,Finkbeiner}.   This effect excludes dark matter with
masses $\la 1-10$ GeV, a much tighter bound than ours (note that such annihilations
also distort the {\it spectrum} of the CMB \cite{MTW,Chluba}, but these bounds are weaker
given present observations).  However, the CMB fluctuation bound only applies to
$s$-wave annihilations, for which $\langle \sigma v \rangle$ does not change
between the dark matter particle freeze-out and the epoch of recombination.  For
$p$-wave annihilations, the annihilation rate at recombination is generally negligible,
and the CMB cannot be used to constrain such models. Therefore, this CMB constraint is applicable
to the model considered in Ref. \cite{Feng} and a dark matter particle with a magnetic dipole moment
\cite{Pospelov,Sigurdson,Gardner,Masso,Fitzpatrick,Cho,Heo1,Heo2,Banks,Barger1,Fortin,
Nobile,Barger2,Heo3}. It is not applicable to the models considered in Refs. \cite{Boehm,Ahn,Hooper}
and a dark matter particle with an electric dipole moment \cite{Pospelov,Sigurdson,Gardner,Masso,Fitzpatrick,Cho,Heo1,Heo2,Banks,Barger1,Fortin,
Nobile,Barger2,Heo3} or an anapole moment \cite{HoScherrer2}, because all of
these models can be $p$-wave dominated.
In these cases our limit provides the better constraint.

In contrast to the CMB constraint, our bounds do not depend on the velocity dependence of the annihilation cross
section and therefore provide a good constraint in the case of $p$-wave annihilations.
Indeed, the values of $N_{eff}$ derived in Refs. \cite{WMAP7B,Dunkley,Keisler} assume
a standard recombination history, undistorted by dark matter annihilation,
so it is unclear how $s$-wave annihilation at the epoch of recombination would affect
the estimated values of $N_{eff}$.  Of course,
the reverse is also true; the bounds derived in Refs. \cite{Padmanabhan,
Mapelli,Zhang,Galli,Hutsi,Finkbeiner} do not take into account the effect we have outlined in this paper.

The bounds presented here can be evaded if the dark matter is asymmetric (see, e.g., Ref.
\cite {LYZ} and references therein).  Also, our bounds will be weakened to the extent that
the dark matter couples to both the electron-photon plasma and to neutrinos.  In fact,
in the extreme opposite limit (coupling to neutrinos only), the $\chi \bar \chi$ annihilation
heats the neutrinos instead of the photons, increasing $N_{eff}$ and providing better
agreement with current observations \cite{Boehm2}.

There is one obvious caveat to the bounds we have derived here.
As noted earlier, the CMB limits on $N_{eff}$ are only in marginal agreement even with
the standard model value for $N_{eff}$.  If future observations show conclusive evidence
that the observed $N_{eff}$ disagrees with the standard model, some mechanism will be required to generate
the additional relativistic degrees of freedom, and this mechanism could also be invoked to erase
the effects of the annihilating dark matter particle. (See, e.g., Ref.
\cite{HoScherrer3}).
Future PLANCK observations should help to resolve this
issue.  More precise observational bounds on $N_{eff}$ would also justify
a more exact treatment of the effect outlined here, going beyond our
simplifying assumption of sudden neutrino decoupling to
a full numerical integration of the equations governing neutrino evolution in the early universe.

\acknowledgments

We thank D. Hooper for helpful discussions.
C.M.H. and R.J.S. were supported in part by the Department of Energy (DE-FG05-85ER40226).

{}


\begin{thebibliography}{99}

\bibitem{Boehm} C. Boehm, D. Hooper, J. Silk, and M. Casse, \prl {\bf 92},
101301 (2004).

\bibitem{Ahn} K. Ahn and E. Komatsu, \prd {\bf 72}, 061301 (2005).

\bibitem{Hooper} D. Hooper and K.M. Zurek, \prd {\bf 77}, 087302 (2008).

\bibitem{Feng} J. L. Feng and J. Kumar, \prl {\bf 101}, 231301 (2008).

\bibitem{Kaplinghat} D. Hooper,
M. Kaplinghat, L. E. Strigari, and K. M. Zurek,
\prd {\bf 76}, 103515 (2007).

\bibitem{KTW}
E. W. Kolb, M. S. Turner, and T. P. Walker,
\prd {\bf 34}, 2197 (1986).

\bibitem{SR}
P. D. Serpico and G. G. Raffelt, \prd {\bf 70}, 043526 (2004).

\bibitem{WMAP7B}
E. Komatsu, et al., Astrophys. J. Suppl. {\bf 192}, 18 (2011).

\bibitem{Dunkley}
J. Dunkley, et al., \apj {\bf 739}, 52 (2011).

\bibitem{Keisler}
R. Keisler, et al., \apj {\bf 743}, 28 (2011).

\bibitem{Dorosh}
A. G. Doroshkevich and M. Yu. Khlopov,
Sov. Astron. Lett. {\bf 9}, 171 (1983).

\bibitem{Dolgov}
A. D. Dolgov, Phys. Rept. {\bf 370}, 333 (2002).

\bibitem{mangano1}
G. Mangano, et al., Nucl. Phys. B {\bf 729}, 221 (2005).

\bibitem{Bashinsky}
S. Bashinsky and U. Seljak, \prd {\bf 69}, 083002 (2004).

\bibitem{Hou}
Z. Hou, et al., arXiv:1104.2333 [astro-ph].

\bibitem{Arch}
M. Archidiacono, E. Calabrese, and A. Melchiorri, \prd {\bf 84}, 123008 (2011).

\bibitem{Weinberg} S. Weinberg, {\it Gravitation and Cosmology},
(New York: Wiley, 1972).

\bibitem{KT} E. W. Kolb and M. S. Turner, {\it The Early Universe}, (New York:
Addison-Wesley, 1990).

\bibitem{Pospelov}
M. Pospelov and T. ter Veldhuis, Phys. Lett. B {\bf 480}, 181 (2000).

\bibitem{Sigurdson}
K. Sigurdson, M. Doran, A. Kurylov, R. R. Caldwell, and M. Kamionkowski,
\prd {\bf 70}, 083501 (2004); erratum, \prd {\bf 73}, 089903 (2006).

\bibitem{Gardner}
S. Gardner, \prd {\bf 79}, 055007 (2009).

\bibitem{Masso}
E. Masso, S. Mohanty, and S. Rao, \prd {\bf 80}, 036009 (2009).

\bibitem{Fitzpatrick}
A. L. Fitzpatrick and K. M. Zurek, \prd {\bf 82}, 075004 (2010).

\bibitem{Cho} W. S. Cho, et al., Phys. Lett. B {\bf 687}, 6 (2010);
erratum, Phys. Lett. B {\bf 694}, 496 (2011).

\bibitem{Heo1} J. H. Heo, Phys. Lett. B {\bf 693}, 255 (2010).

\bibitem{Heo2} J. H. Heo, Phys. Lett. B {\bf 702}, 205 (2011).

\bibitem{Banks}
T. Banks, J. -F. Fortin, and S. Thomas, arXiv:1007.5515 [hep-ph].

\bibitem{Barger1}
V. Barger, W. -Y. Keung, and D. Marfatia, Phys. Lett. B {\bf 696}, 74 (2011).

\bibitem{Fortin}
J. -F. Fortin and T. M. P. Tait, \prd {\bf 85}, 063506 (2012).

\bibitem{Nobile}
E. Del Nobile, et al., JCAP {\bf 1208}, 010 (2012).

\bibitem{Barger2}
V. Barger, W. -Y. Keung, D. Marfatia, and P. -Y. Tseng, Phys.\ Lett.\ B {\bf 717}, 219 (2012).

\bibitem{Heo3}
J. H. Heo and C. S. Kim, arXiv:1207.1341 [astro-ph].

\bibitem{HoScherrer2}
C. M. Ho and R. J. Scherrer, arXiv:1211.0503 [hep-ph]. 

\bibitem{Enqvist}
K. Enqvist, K. Kainulainen, and V. Semikoz, Nucl. Phys. B {\bf 374},
392 (1992).

\bibitem{Hannestad}
S. Hannestad, \prd {\bf 65}, 083006 (2002).

\bibitem{HoScherrer3}
C. M. Ho and R. J. Scherrer, arXiv:1212.1689 [hep-ph].

\bibitem{Beacom}
J. F. Beacom and H. Yuksel, \prl {\bf 97}, 071102 (2006).

\bibitem{Steigman}
G. Steigman, Ann. Rev. Nucl. Part. Sci. {\bf 57}, 463 (2007).

\bibitem{Padmanabhan}
N. Padmanabhan and D. P. Finkbeiner, \prd {\bf 72}, 023508 (2005).

\bibitem{Mapelli}
M. Mapelli, A. Ferrara, and E. Pierpaoli, Mon. Not. R. Astr. Soc.
{\bf 369}, 1719 (2006).

\bibitem{Zhang} L. Zhang, X. Chen, Y. -A. Lei, and Z. -g. Si,
\prd {\bf 74}, 103519 (2006).

\bibitem{Galli}
S. Galli, F. Iocco, G. Bertone, and A. Melchiorri,
\prd {\bf 84}, 027302 (2011).

\bibitem{Hutsi}
G. Hutsi, J. Chluba, A. Hektor, and M. Raidal,
Astron.\ Astrophys.\  {\bf 535}, A26 (2011).

\bibitem{Finkbeiner}
D. P. Finkbeiner, S. Galli, T. Lin, and T. R. Slatyer,
Phys.\ Rev.\ D {\bf 85}, 043522 (2012).

\bibitem{MTW}
P. McDonald, R. J. Scherrer, and T. P. Walker,
\prd {\bf 63}, 023001 (2001).

\bibitem{Chluba}
J. Chluba and R. A. Sunyaev, Mon. Not. R. Astr. Soc. {\bf 419}, 1294 (2012).

\bibitem{LYZ}
T. Lin, H. -B. Yu, and K. M. Zurek, \prd {\bf 85} 063503 (2012).

\bibitem{Boehm2}
C. Boehm, M. J. Dolan, and C. McCabe, arXiv:1207.0497 [astro-ph].

\end{thebibliography}
\end{document}